\documentclass[prl,twocolumn,longbibliography]{revtex4-1}

\usepackage{color}
\usepackage{amsmath}
\usepackage{xspace}
\usepackage{amsfonts}
\usepackage{tikz}
\usepackage[squaren,Gray]{SIunits}

\usepackage{amsthm}
\theoremstyle{plain}

\theoremstyle{remark}

\newcommand{\dd}{\mathrm{d}}
\newcommand{\superop}[1]{\ensuremath{\mathcal{#1}}\xspace}
\newcommand{\op}[1]{\ensuremath{\mathbf{#1}}\xspace}
\renewcommand{\vec}[1]{\ensuremath{\mathbf{#1}}\xspace}
\newcommand{\avg}[1]{\ensuremath{\langle #1\rangle}\xspace}

\usepackage{caption}

\usepackage{soul}
\newcommand*{\revadd}[1]{\textcolor{red}{#1}}
\newcommand*{\revrem}[1]{\textcolor{red}{\textst{#1}}}

\renewcommand{\revadd}[1]{#1}
\renewcommand{\revrem}[1]{}

\begin{document}

\title{Molecular hydrodynamics from memory kernels}
\author{Dominika Lesnicki}
\affiliation{\'Ecole Normale Sup\'erieure, PSL Research University,
D\'epartement de Chimie, Sorbonne Universit\'es -- UPMC Univ Paris 06, CNRS
UMR 8640 PASTEUR, 24 rue Lhomond, 75005 Paris, France}
\author{Antoine Carof}
\affiliation{Sorbonne Universit\'es, UPMC Univ Paris 06, CNRS, Laboratoire PHENIX, Case 51, 4 place Jussieu, F-75005 Paris, France}
\author{Benjamin Rotenberg}
\affiliation{Sorbonne Universit\'es, UPMC Univ Paris 06, CNRS, Laboratoire PHENIX, Case 51, 4 place Jussieu, F-75005 Paris, France}
\author{Rodolphe Vuilleumier}
\affiliation{\'Ecole Normale Sup\'erieure, PSL Research University,
D\'epartement de Chimie, Sorbonne Universit\'es -- UPMC Univ Paris 06, CNRS
UMR 8640 PASTEUR, 24 rue Lhomond, 75005 Paris, France}

\begin{abstract}

The memory kernel for a tagged particle in a fluid, computed from molecular dynamics simulations, decays algebraically as $t^{-3/2}$. We show how the hydrodynamic Basset-Boussinesq force naturally emerges from this long-time tail and generalize the concept of hydrodynamic added mass. This mass term is negative in the present case of a molecular solute, at odds with incompressible hydrodynamics predictions. We finally discuss the various contributions to the friction, the associated time scales and the cross-over between the molecular and hydrodynamic regimes upon increasing the solute radius.

\end{abstract}
\maketitle

The Brownian motion of a particle in a fluid finds its origin in the fluctuating
force exerted by the solvent molecules on the solute. It has long been known
that the canonical description of this random force by a Gaussian Markov process
is only valid in limiting cases. Even in the limit where the solute is much
heavier than the solvent particles, for which
multiple time-scale analysis allows to recover the Smoluchowski equation for
diffusion~\cite{Bocquet1998}, non-Markovian effects are expected when the mass
density ratio is close to unity~\cite{Bocquet1997} -- a situation which is rather
the rule than the exception \textit{e.g.} in colloidal suspensions.
These non-Markovian effects arise because of momentum conservation, leading to
slow hydrodynamic modes that manifest themselves as long-time tails in the
velocity autocorrelation function
(VACF)~\cite{AlderWainwright1970,HansenBook,Oppenheim1975,OuldKaddour2000}. Recent 
experiments have demonstrated that the force exerted by the bath includes a deterministic
component~\cite{Franosch2011}, well described for large colloidal spheres by the
Basset-Boussinesq (BB) hydrodynamic force~\cite{Boussinesq1903,ChowHermans1972}:
\begin{align}
\nonumber
\label{eq:BB}
\vec{F}_{BB}(t)&=
-6\pi\eta R \vec{v}(t) 
- {2 \over 3}\pi R^3\rho_0\dot{\vec{v}}(t)
\\
& - 6R^2\sqrt{\pi \rho_0 \eta}\int_0^t(t-u)^{-{1 \over 2}}\dot{\vec{v}}(u)\,\dd u
\;,
\end{align}
where $R$ is the sphere radius, $\eta$ the solvent viscosity and $\rho_0$ its
mass density. The first term is the usual Stokes friction. The other two account
for the inertia of the displaced fluid and involve a finite added mass 
$m_0^{BB}={2 \over3}\pi R^3\rho_0$ and a viscosity-dependent retarded component
describing the transient effects of momentum diffusion in the solvent.

While continuous descriptions of steady-state flows appear to hold down to the
nanoscale~\cite{BocquetBarrat1993,BocquetBarrat1994,BocquetBarrat2007}, 
possibly at the price of adapting the hydrodynamic radius $R$ or the boundary
conditions~\cite{Skinner2003}, their validity for the transient regimes
should be questioned. The implicit assumption of a separation of time scales 
between the solvent and solute dynamics, which holds a priori for colloidal 
particles~\cite{Padding2006}, is expected to break down with smaller solutes 
such as nanoparticles or biomolecules.

Here we address the fundamental questions that arise when 
approaching the regime of molecular solutes by computing directly from 
Molecular Dynamics (MD) simulations the memory kernel and the random noise of the 
Generalized Langevin Equation (GLE). A novel algorithm based on the
Mori-Zwanzig formalism with high numerical stability allows us to explore
long time scales for the first time.
We consider the extreme case of a tagged particle (identical masses and sizes) in
a pure supercritical fluid. 

By examining the long-time behaviour of the memory kernel,
we demonstrate the generality of the functional form of Equation~\ref{eq:BB}
beyond pure hydrodynamic descriptions
and discuss its interpretation as the
time-dependent force exerted by the solvent on the solute at thermal
equilibrium. Importantly, we show how to define and compute a mass
from the memory kernel itself. 
This generalisation from the microscopic dynamics
correctly describes the numerical results for the VACF almost down
to the ballistic time scale and provides insights into the emergence of the
hydrodynamic behaviour for larger solutes, bridging the gap between the 
solvent and colloidal time scales.

In the Zwanzig-Mori formalism~\cite{ZwanzigBook,Mori1965,Grabert1982}, the velocity $\op{v}(t)$ of a tagged particle of mass $m$ in a fluid follows the generalized Langevin equation
\begin{equation}
\label{eq:gle}
m\frac{\dd \op{v}}{\dd t}(t)=-\int_0^t K(u) \op{v}(t-u)  \,\dd u + \op{R}(t),
\end{equation}
where $K(u)$ is the memory kernel and $\op{R}(t)$ the so-called random force, that are obtained from the true force $\op{F}$ acting on the tagged particle using the projection operator technique and defined as
\begin{equation}
\label{eq:memory}
K(u) = \frac{\avg{{\op{F}}\, e^{i(1-\superop{P})\superop{L}u}\op{F}}}{k_BT}
 = \frac{\avg{{\op{F}}\,\op{R}(u)}}{k_BT}
\end{equation}
with $k_B$ the Boltzmann constant and $T$ the temperature, and
$\op{R}(t)= e^{i(1-\superop{P})\superop{L}t}  \op{F}$.
In these equations, $i\superop{L}$ is the Liouvillian operator corresponding to the unperturbed dynamics and $\superop{P}$ is the \revadd{Mori projection operator along the velocity \op{v}}, acting on an observable $\op{A}$ as
$\superop{P}\op{A}=\frac{\avg{\op{v}\op{A}}}{\avg{\op{v}^2}}\op{v}$.
Throughout the paper, $\avg{\cdot}$ denotes the canonical equilibrium average at temperature $T$. The force $\op{F}$ is propagated using the orthogonal dynamics $e^{i(1-\superop{P})\superop{L}u}$ instead of the normal dynamics to obtain the Zwanzig-Mori memory kernel. 

The auto-correlation function of this projected force, or noise, 
$\avg{{\op{F}}\, e^{i(1-\superop{P})\superop{L}u}\op{F}}$ differs significantly
from the auto-correlation function of the force $\avg{\op{F}\,\op{F}(u)}$. In
particular, for a periodic system
the latter integrates to zero whereas the former integrates to the
friction $\xi$. This property of the projected force is a form of the Einstein relation since the friction is
related to the diffusion constant $D$ of the tagged particle by
$D=\frac{k_BT}{\xi}$. However, extracting the projected force correlation
function, or the kernel, from MD simulations is a difficult task. 

We have recently introduced two practical schemes to compute such
properties for generic observables from MD trajectories~\cite{Carof2014}. 
These algorithms are only accurate to first order in the MD timestep $\delta t$
-- thus preventing their use to investigate the long time behaviour. 
Here we employ a novel algorithm~\cite{SuppMat}\nocite{KuboBook,NIST,Malbrunot1983}, which provides
second order accuracy at virtually no additional computational cost, to study
the memory kernel for diffusion in a Lennard-Jones (LJ) fluid. We consider a
system of 10$^4$ LJ particles at a reduced density $\rho^*=\rho \sigma^3 = 0.5$ and reduced temperature $T^*=k_BT/\epsilon=1.5$
with $\sigma$ and $\epsilon$ the LJ diameter and energy, respectively,
\textit{i.e.} at the critical density and slightly above the critical temperature. 
Newton's equations of motion are solved using the
velocity Verlet algorithm and cubic periodic boundary
conditions. Interactions are computed using a cut-off radius $r_c=3\sigma$. The
system is first equilibrated at the target temperature during 230.41~$t^*$ by
performing MD with a timestep of 9.2$\times10^{-4}$~$t^*$, 
in the NVT ensemble using Langevin thermostat with
a time constant of 0.92~$t^*$. All properties are then determined from a 230.41~$t^*$
trajectory with a timestep of 4.6$\times10^{-4}$~$t^*$ in the NVE ensemble generated with the
DLPOLY~\cite{DLPOLY} simulation package and 
block averages were taken over trajectory segments of one tenth of the total trajectory.

The novel second order algorithm presents remarkable long time stability and allows to
investigate time scales much beyond $\sim t^*$. This is demonstrated in the inset of the
figure~\ref{fig:acfs}, which displays the running time integral of the noise auto-correlation
function (NACF). 
From the plateau of the NACF (Fig.~\ref{fig:acfs}), 
we obtain $\xi\sim 4.5\pm 0.1$ LJ units, in excellent
agreement with the Einstein relation ($k_BT/D\sim4.4$ LJ units).
In contrast, the running time integral of the unprojected force auto-correlation function (FACF) tends to zero as expected. 
\begin{center}
\captionsetup{type=figure}
 \includegraphics*[width=1.0\linewidth]{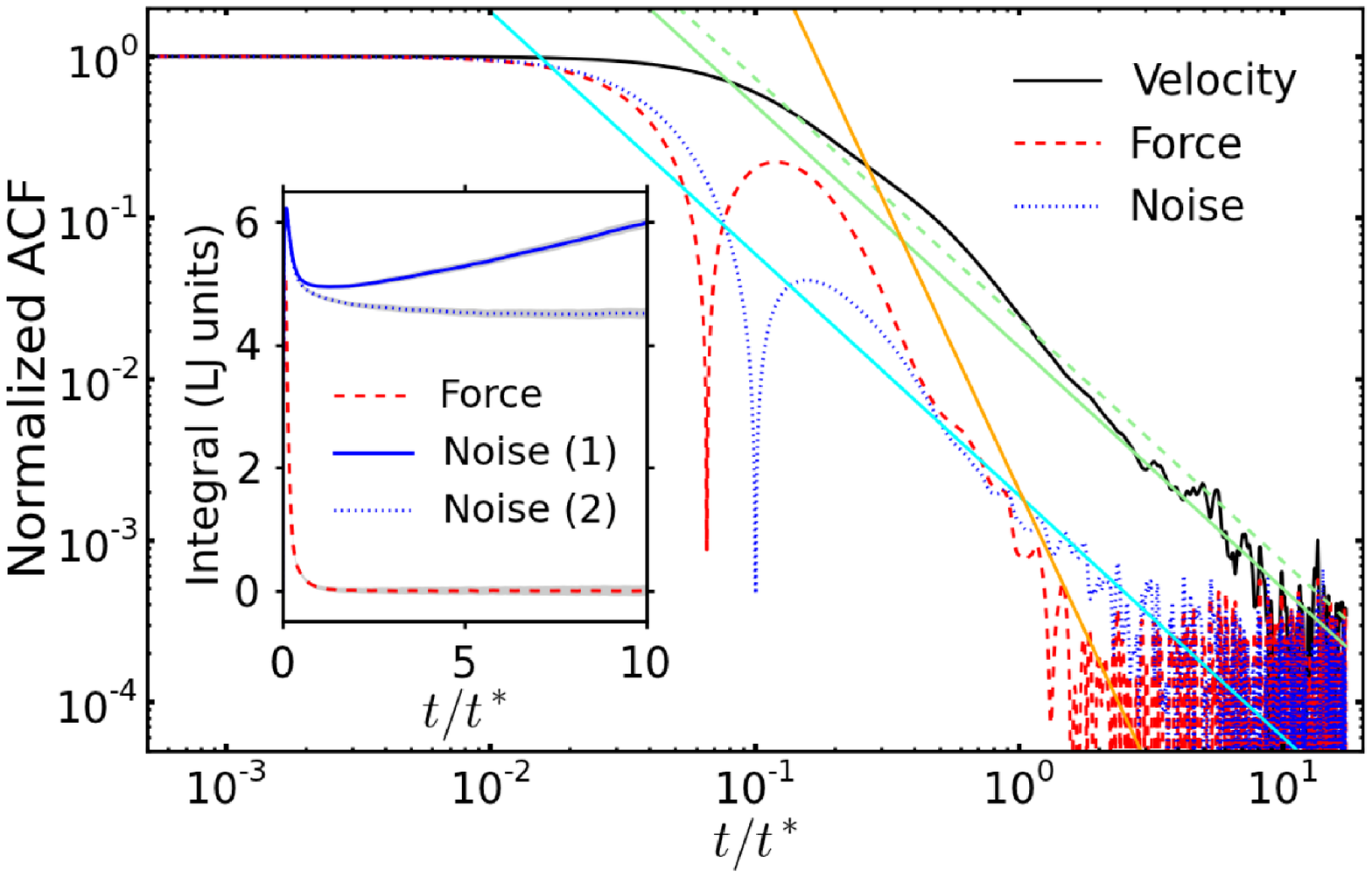}
\caption{\label{fig:acfs}\small{Absolute value or the normalized velocity (solid black), force (dashed red) and noise (dotted
blue) auto-correlation functions, as a function of time in Lennard-Jones units.
Molecular simulation results are compared to the hydrodynamic scalings: The green lines correspond to Eq.~\ref{eq:zoft} without (solid) and neglecting (dashed) the diffusion coefficient, the orange and cyan lines to Eqs.~\ref{eq:facf} and~\ref{eq:nacf}, respectively. The inset displays the corresponding integrals for the force (dashed red), which converges to zero as expected, and noise: While it diverges with the first order algorithm (solid blue line), it converges to the friction $\xi$ with the novel second order one (dotted blue line). Grey area indicate the error bars.}}
\end{center}

Figure~\ref{fig:acfs} then shows the long-time behaviour of the normalized
velocity, force and noise autocorrelation functions. Hydrodynamic
and mode coupling theories predict that the VACF decays at long times as
\begin{equation}
\label{eq:zoft}
Z(t)={1\over3}\avg{\op{v}\cdot\op{v}(t)}\sim \frac{2k_BT}{3\rho m} [4\pi (D+\nu)t]^{-{3 \over 2}}
\; ,
\end{equation}
where $\nu={\eta \over \rho m}$ is the kinematic viscosity, with $\eta$ the
fluid viscosity and $m$ the particle mass. The diffusion constant $D$ is often
omitted in this long time tail, however it is necessary to reproduce our
numerical result as can be seen from figure~\ref{fig:acfs}. This term is due to the
diffusion of the particle simultaneously with the momentum transfer in the
fluid~\cite{Mazur1970,Donev2013}.  The FACF is the second-order
derivative of the VACF and should decay in the same limit as:
\begin{equation}
\label{eq:facf}
\langle F(t)F(0)\rangle = \frac{{\rm d}^2}{{\rm d}t^2}Z(t) 
\sim t^{-{7 \over 2}}
\; .
\end{equation}
This is indeed the case as shown in figure~\ref{fig:acfs}.
In contrast, the NACF, which is nothing but the memory kernel $K$, decays much
more slowly than the FACF, following the same $t^{-{3 \over 2}}$ scaling as the
VACF. In fact, such a scaling is not unexpected: Corngold indeed showed from the relation between the Laplace transforms of $Z(t)$ and $K(t)$
that under rather mild conditions for the VACF, the memory kernel defined by
Eq.~\ref{eq:gle} should decay as~\cite{Corngold1972}
$K(t) \sim -\frac{\xi^2}{k_BT}Z(t)$
leading to
\begin{equation}
\label{eq:nacf}
K(t) \sim -\frac{2\xi^2}{3\rho m} [4\pi (D+\nu)t]^{-{3 \over 2}}
\; ,
\end{equation}
from the asymptotic behavior of $Z(t)$.
As can be seen in figure~\ref{fig:acfs}, this prediction is indeed
satisfied by the memory kernel determined from MD.
This scaling is also consistent with the low frequency limit of the 
hydrodynamic memory kernel corresponding to Eq.~\ref{eq:BB} (see below). It has then been
observed experimentally for colloidal particles where this limit applies~\cite{Franosch2011}. Our
results confirm for the first time that this scaling also holds for the
diffusion of microscopic particles. 

From the decay of the memory kernel at long times, the development of the Laplace transform of the friction kernel is
\begin{equation}
\label{eq:Kofs}
\tilde{K}(s)=\xi + \alpha s^{1/2} + m_0s + o(s), 
\end{equation}
with:
\begin{equation}
\alpha\pi^{-1/2}=\frac{4}{3}\xi^2 \frac{1}{\rho m [4\pi(\nu+D)]^{3/2}}
\;,
\end{equation}
where we have introduced a mass defined by:
\begin{equation}
m_0 \equiv-\int_0^{+\infty} \left(K(t)+\frac{1}{2}\alpha\pi^{-1/2} t^{-3/2}\right)t\, \dd t
\; ,
\label{eq:mzero}
\end{equation}
under the assumption that
$K(t)+\frac{1}{2}\alpha\pi^{-1/2} t^{-3/2}$ decreases to zero faster than
$t^{-2}$, and where an integration by parts was used for the second equality. 
\revadd{Note that while the speed of convergence depends on higher order terms in the expansion Eq.~\ref{eq:Kofs}, the value of $m_0$ defined by Eq.~\ref{eq:mzero} does not.}
Figure~\ref{fig:mzero} shows the running integral associated with the
definition of this mass. The observed plateau demonstrates the convergence
of the integral and thus validates the above assumption in the present case.
Eq.~\ref{eq:mzero} therefore provides the first definition of the mass term from the microscopic dynamics.

\begin{center}
 \captionsetup{type=figure}
 \includegraphics*[width=1.0\linewidth]{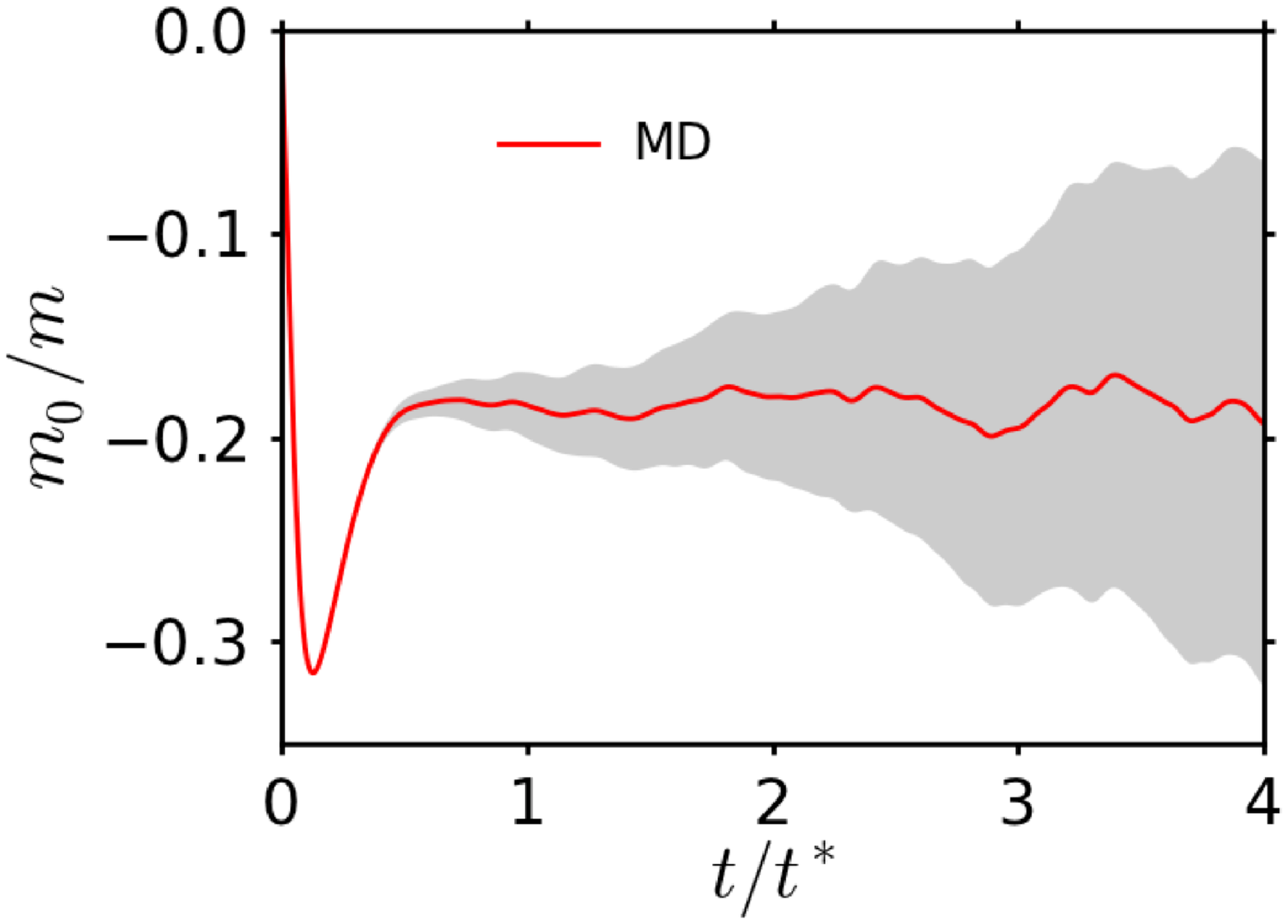}
 \caption{\label{fig:mzero}\small{Running integral defining the mass $m_0$, normalized by the
mass $m$ of the Lennard-Jones particle (see Eq.~\ref{eq:mzero}).
Grey area indicate the error bars.
}}
\end{center}

Surprisingly, this mass term is negative, with a value of $m_0\sim-0.18~m$
-- in contradiction with the incompressible hydrodynamic prediction \revadd{for the added mass}
$m_0^{BB}={2\over3}\pi R^3 \rho_0$~\revadd{\cite{Donev2013,Maxey1983}}.
This observation can be interpreted as follows, by analyzing the various
contributions to the kernel $K$. At short times, $K$ is dominated by
short-range collisions between the solute and the solvent and can be
approximated by an exponential decay $K(t) = {\xi_E\over\tau_0}e^{-t/\tau_0}$,
with $\xi_E$ the Enskog friction~\cite{HansenBook} and $\tau_0$
the characteristic time for the decay of the FACF ($\tau_0\sim0.05~t^*$
in the present case). The subscript $0$ indicates that this time corresponds to
the collisions between the solvent molecules and the solute, rather than to a
time scale associated with the decay of the solute VACF. This collisional
component of the kernel contributes to the mass defined in Eq.~\ref{eq:mzero} 
as a \emph{negative} term $m_0^E=-\xi_E\tau_0$. Computing the Enskog
friction~\cite{Bocquet1994} for a solute of size $\sigma$, we get $\xi_E\approx5.8$.
This value is consistent with the maximum of the time-dependent friction
in figure~\ref{fig:acfs} (see below) and results for the mass to a contribution
$m_0^E\approx-0.29~m$. 

Other mechanisms come into play on times scales longer than $\tau_0$.
Indeed, momentum transfer from the solute to the solvent
includes a transient regime giving rise to a positive contribution to 
the mass term (here over a time $\tau_m\sim0.5~t^*$, 
as can be seen in Fig.~\ref{fig:mzero}) and eventually becomes diffusive, leading to the 
retarded force and to a decrease in the friction 
(see Fig.~\ref{fig:acfs}): The solvent backflow tends to drag the solute in the direction of its initial velocity, 
\textit{i.e.} contributes negatively to the friction.
Assuming that this component of the mass term is well described
by the hydrodynamic result despite the molecular size of the solute,
we obtain a total mass $m_0=m_0^E+m_0^{BB}\approx-0.16~m$, which is
in good agreement with the MD result considering the strong assumptions 
involved (validity of the Enskog result at high packing fraction and
hydrodynamic model of the mass), and conforts our interpretation of the two competing contributions to
the mass term. 

\revadd{We now consider the ensemble-average velocity $\bar{v}$ obtained over an ensemble of identical systems initially in equilibrium and put out of equilibrium at time $t=0$ by a time-dependent applied force ${f}_\epsilon(t)$, identical to all replicas of the system.}
We \revrem{further} show in Supplementary Material that the evolution of the
ensemble average velocity $\bar{v}$ is given by the same kernel as the GLE for
the microscopic velocity with the random force replaced by the applied
force~\cite{SuppMat}.
For slowly varying forces, the ensemble-average velocity also varies slowly and
we can consider the $s\to0$ limit in $\tilde{K}(s)$. The first three terms of Eq.~\ref{eq:Kofs} correspond to an evolution of $\bar{v}$, 
according to:
\begin{align}
\label{eq:BBmol}
m\dot{\bar{v}}(t) &= 
 f_{\epsilon}(t)
-\xi\bar{v}(t) 
- m_0\dot{\bar{v}}(t) 
\nonumber \\ & 
- \alpha\pi^{-1/2}\int_0^t(t-u)^{-{1 \over 2}}\dot{\bar{v}}(u)\,\dd u
\;,
\end{align}
for a system put out of equilibrium from $t=0$ 
by a slowly-varying infinitesimal force $f_\epsilon(t)$. 
\revrem{The ensemble-average is made over initial conditions taken at thermal equilibrium for the same time-dependent external force $f_\epsilon(t)$.}
There is no hypothesis of separation of time scales between slow and fast degrees of
freedom of the system in this equation, its meaning is that of a slowly varying
response to a slow perturbation~\cite{HynesTransient1973}.
This evolution provides a generalisation of the Basset-Boussinesq Eq.~\ref{eq:BB} to
arbitrary solutes satisfying only the above generic assumptions 
on the long-time behaviour of the corresponding memory kernel.

\begin{center}
\captionsetup{type=figure}
 \includegraphics*[width=1.0\linewidth]{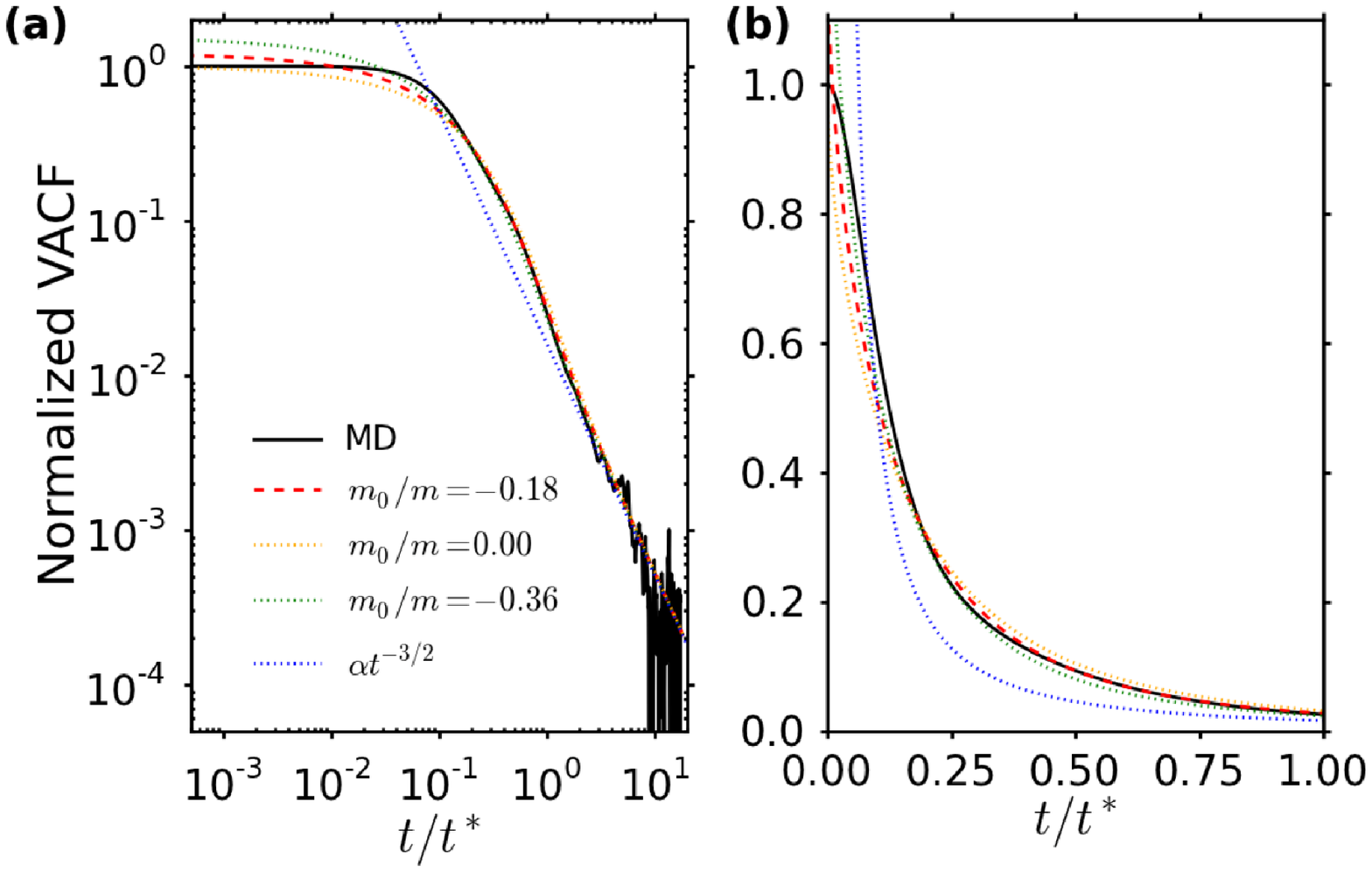}
 \caption{\label{fig:hydro}\small{Normalized velocity auto-correlation function (a: logarithmic scale,
b: linear scale for short times). The molecular simulation
results (black solid line) are compared to the long-time scaling
(Eq.~\ref{eq:zoft}, blue dotted line) and the analytical result 
of Ref.~\cite{ChowHermans1972} corresponding to the memory kernel
Eq.~\ref{eq:Kofs} with three values of the mass $m_0$.
}}
\end{center}

Following the method of Chow and Hermans for
the VACF of a particle subject to a BB force~\cite{ChowHermans1972}, we express analytically the VACF of the solute subject to the force Eq.~\ref{eq:BBmol}
and compare it to the simulation results in Figure~\ref{fig:hydro}, both in logarithmic
and linear scales. The agreement is excellent down to relatively short times
(less than 0.5$t^*$), without any adjustable parameter. This further
demonstrates the relevance of the above definition of the mass term from the
memory kernel (\textit{i.e.} not from the solute geometry and hydrodynamic
properties of the solvent). 
Note, however, that truncating the memory kernel to the first three terms
of the low frequency expansion Eq.~\ref{eq:Kofs} leads to some limitations for the
description of the short-time behaviour, such as an incorrect initial value of the VACF, namely 
$k_BT/(m+m_0)$ instead of $k_BT/m$~\cite{ChowHermans1972}.

A negative contribution to the mass term can also be derived
from the hydrodynamics of compressible fluids, involving the time $R/c$ 
it takes for sound waves to propagate over the particle radius
-- confirming the role of retardation effects in this negative contribution. 
However, introducing compressibility in continuum hydrodynamics~\cite{Chow1973,Chakraborty2011} does not
improve the prediction for the VACF, even with an effective
hydrodynamic radius adjusted to reproduce the calculated friction
(see~\cite{SuppMat}), because it does not capture molecular scale effects.

\begin{center}
\captionsetup{type=figure}
 \includegraphics*[width=1.0\linewidth]{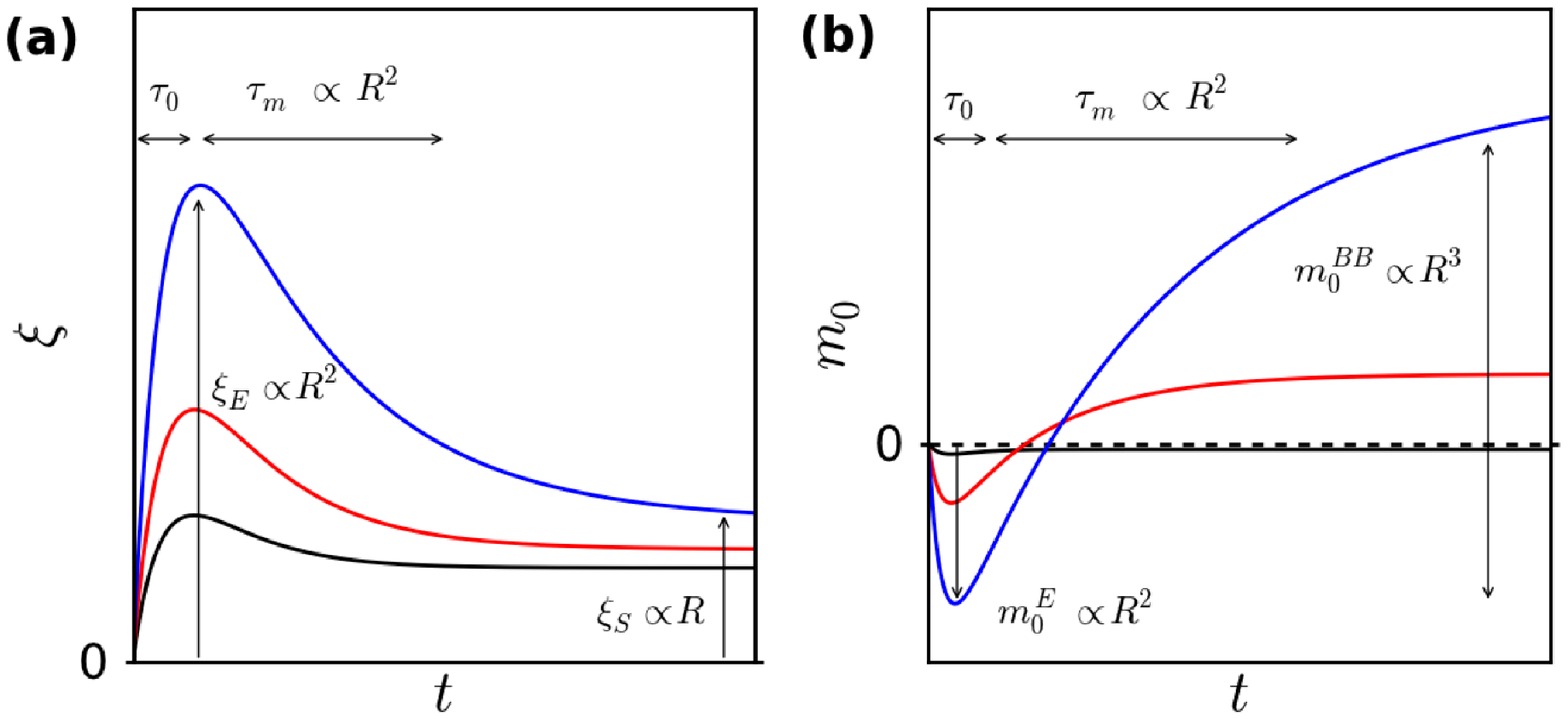}
\caption{\label{fig:evolradius}{Schematic evolution of the friction (a) and mass (b)
with increasing solute radius $R$ for a spherical density-matched solute.
The curves correspond to increasing radius from black to red to blue.
}}
\end{center}

Finally, let us consider the implications of the present work for larger
solutes. We consider here spherical solutes with a density equal to that of the
solvent, which is the most common experimental situation of density-matched
colloidal suspensions, with a mass $M=\frac{4\pi R^3}{3}\rho_0$.
The following discussion is illustrated in Figure~\ref{fig:evolradius}.
For a large particle ($R\gg\sigma/2$ and $M\gg m$), the Enskog friction
$\xi_E\propto R^2$ and the negative Enskog contribution $-\xi_E\tau_0$
to the mass (where $\tau_0$ only weakly depends on $R$)
dominate at short times. The Stokes friction $\xi_S\propto R$
and the BB hydrodynamic mass $m_0^{BB}\propto R^3$ are
recovered over a time $\tau_m\propto R^2/\nu$. 
While in the present case this analysis neglects molecular features, $\tau_m$ provides the correct order of
magnitude for the time over which the integral defining the mass converges
(see Fig.~\ref{fig:mzero}). 

The long time-tail of the memory kernel
bridges microscopic dynamics with continuum hydrodynamics as 
it gives rise to a force entering in the evolution equation of the tagged
velocity similar to the BB hydrodynamic force. 
The memory kernel further allows for the first microscopic definition 
of the mass present in this evolution equation. This mass is 
found to be negative for a solute identical to solvent particles and is
related to the retardation of the friction force.
Extracting the mass term directly
from MD simulations paves the way to the study of isotopic effects. It
can also be used to quantify in a well-defined way the number of molecules
brought along ions during transport 
or to interpret the peculiar behaviour of the friction on alcanes as a function 
of chain length~\cite{Lee2003,Falk2015}. In particular, it provides a
microscopic route to model acoustophoresis~\cite{Debye1933,Bernard1995}
or electro-osmotic effects~\cite{Marry2003,Rotenberg2013}. Finally, the novel
algorithm introduced here could be used to compute projected correlation functions
of other observables and to investigate the properties of the corresponding GLE.

The authors are endebted to Jean-Pierre Hansen and Lyd\'eric Bocquet for very
fruitful discussions and for critical reading of the manuscript.


\bibliography{article}

\end{document}